\documentstyle[12pt]{article}
\newcommand{\Section}[1]{\section{#1}}

\def\be{\begin{equation}}
\def\ee{\end{equation}}
\def\bea{\begin{eqnarray}}
\def\eea{\end{eqnarray}}
\def\ba{\begin{array}}
\def\ea{\end{array}}
\def\s{\sum\limits}
\def\tp{\otimes}
\def\L{\Lambda}

\def\t{\theta}
\def\d{\partial}
\def\tb{{\bar{\theta}}}
\def\fb{{\bar{f}}}
\def\zb{{\bar{z}}}
\def\f{\frac}
\def\a{\alpha}
\def\b{\beta}
\def\g{\gamma}
\def\o{\omega}
\def\hs{\hspace}
\def\ff{\f{1}{2}}
\def\p{\phi}

\def\rar{\rightarrow }
\def\os2{$osp(2\mid 2)$}
\def\os{osp(2|2)}
\def\uos2{$U_q(osp(2|2))$}
\def\PP{<\Phi(z_1,\t_1)\Phi(z_2,\t_2)>}
\def\PPP{<\Phi(z_1,\t_1)\Phi(z_2,\t_2)
         \Phi(z_3,\t_3,)>}
\begin{document}

\begin{titlepage}
\vspace{-10mm}
\vspace{12pt}
\begin{center}
\begin{Large}
             {\bf $N=2$ Super-Conformal Filed Theory on the Basis of
$\os$}
\end{Large}

\
\

                          {\bf A.
Shafiekhani${}^\dagger${\footnote {e-mail: ashafie@theory.ipm.ac.ir}}
and W.-S. Chung${}^*$}\\
\vspace{12pt} {\it
\vspace{8pt}
${^\dagger}$Institute for Studies in Theoretical Physics and Mathematics\\
             P.O.Box: 19395-5746, Tehran, Iran,\\

                ${^*}$Theory Group, Dept. of Physics\\
                College of Natural Sciences,\\
                Gyeongsang National University, Jinju, 660.701, Korea\\
}

\

\end{center}
\abstract{

Using a unified and systematic scheme,
the free field realization of irreducible representations of $\os$ is
constructed. By using these realization, the
correlation functions of $N=2$ super-conformal model based on
$\os$
symmetry and free field representation of ${\widehat{\os}}$ generators
are calculated.
Free field representation of currents are used to determine the stress-energy
tensor and the central charge of the model.
\vfill
}

\end{titlepage}
{\Section {Introduction}}

There are many evidences that 2-dimensional $N=2$ super-conformal field
theories are very important in
string theory \cite{string}, disordered systems at criticality \cite{mas,ckt},
topological field theories \cite{topo}, integrable models \cite{morz},
mirror symmetry \cite{gw} and
quantum $W$-gravity \cite{grav}. Such interests are enough to motivate one
to study different aspects of $N=2$ in 2-dimension.

A problem of much physical interest is the calculation of correlation functions,
using super-conformal Ward identities.

From the other side
free field representations of Lie (super)algebras \cite{mor,az,wa} are of interest
because, they have many applications in (super)conformal field
theories \cite{cft}, inverse problems and integrable systems \cite{andre,frt},
and statistical mechanics \cite{sm,mas,ckt}.
In any conformal or super-conformal field theory
all the properties of the
theory can be encoded in its OPE's. The main point in any OPE calculation is to
use as much symmetry as possible for simplification. If somehow one can get free
field representation of
currents \cite{mor} the OPE calculation will be straight forward.

In the case of super-symmetric conformal field theories, it is often useful to
work in a super-space formalism.  Using super-space we will be able to find
coherent states and the differential realization for generators of conformal
symmetry. Such a realization can be used to find free field representations of
 current algebra in level zero, then using Feigin and Frankel method
\cite{fefr} to extend it to any arbitrary level. On the other hand we will be
able to use differential realization in Ward identities to calculate
correlation functions of the conformal field theory based on $\os$
symmetries.

In this paper
we give a very simple way of calculation for their free field representation,
which is extendable to all other symmetric groups.

Our aim in this paper is to apply the unified and systematic scheme given in
ref. \cite{az,wa}, for $\os$ and use these realization
to calculate correlation functions of $N=2$ super-conformal models
based on $\os$ symmetry group.

The structure of this paper is as follows: In section 2, we construct
the differential realization of $\os$ for all its irreducible
representations.
In section 3, we will use the results of
section 2 to calculate two and three point functions of super-conformal 
field theory based on $\os$ symmetry group. In section 4, we will use
the differential realization of $\os$ to calculate free field 
representation of ${\widehat {osp(2|2)}}$
algebra, and stress-energy tensor of the theory.

{\Section {Differential realization of $\os$}}
The non-compact super-group $OSP(2|2)$ is generated by
eight elements \cite{n1,m,bsh}.
The four even generators, $X_\pm$, $H$ and $B$, close under commutation and
generate the subgroup of
$sl(2)\times U(1)$ and with four others odd generators $V_\pm$ and $W_\pm$,
satisfy the following (graded)commutation relations

\be \label{osc}
\ba{lll}
[H, X_\pm]=\pm X_\pm & [X_+, X_-]=-2H & [B, X_\pm]=0\cr
[B, H]=0 & [H, V_\pm]=\pm\ff V_\pm & [H, W_\pm]=\pm\ff W_\pm \cr
[X_\pm, V_\pm]=0 & [X_\pm, W_\pm]=0 &[X_\pm, V_\mp]=\mp V_\pm \cr
[X_\pm, W_\mp]=\mp W_\pm & [B, V_\pm]=\ff V_\pm &[B,W_\pm]=-\ff W_\pm\cr
\{ V_\pm, V_\pm \} =0 &\{ V_\pm, V_\mp \} =0 & \{ W_\pm, W_\pm \} =0 \cr
\{ W_\pm, W_\mp \} =0& \{ V_\pm, W_\pm \} =X_\pm & \{ V_\pm, W_\mp \} =H\mp B
\ea
\ee
Let us define $<b,h;-2b,-2h,-2h|:=<-2b,-2h|$ as the highest weight
of an irreducible representation of $\os$, where
in left hand side the first two quantum numbers label the $\os$ representation,
while the next three labels are $U(1)$ quantum number, the $sl(2)$ quantum
number and the corresponding third component, respectively.
Such a highest weight is annihilated by all raising generators:
\be\ba{ll}
<-2b,-2h|X_+=0,&<-2b,-2h|V_+=0,\cr
<-2b,-2h|H=<-2b,-2h|(-2h),&<-2b,-2h|W_+=0, \cr
<-2b,-2h|B=<-2b,-2h|(-2b)&.
\ea\ee
The relevant coherent state is as follows:
\be
<-2b,-2h|e^{(z-\ff\tb\t)X_-}e^{\tb W_-}e^{\t V_-}:=<z,\t,\tb|
\ee
which $z$ is complex variable and $\t$ and $\tb$ are Grassmanian variables
\cite{bsh}.

By the similar method as developed in \cite{az,wa}, by acting the group 
generators on the above coherent states and using the above 
(graded)commutation relations,
the differential realization of $\os$ for left action will be
as follows:
\be\ba{c}\label{lrz}
V_-=\d_{\tb}+\ff\t\d_z\cr
W_-=\d_{\t}+\ff\tb\d_z\cr
V_+=z\d_{\tb}+\ff z\t\d_z+\ff\t\tb\d_{\tb}-2\t(h-b)\cr
W_+=z\d_{\t}+\ff z\tb\d_z-\ff\t\tb\d_{\t}-2\tb (h+b)\cr
H=\ff\t\d_\t+\ff\tb\d_{\tb}+z\d_z-2h\cr
B=\ff\t\d_\t-\ff\tb\d_{\tb}-2b\cr
X_-=\d_z\cr
X_+=z^2\d_z+z\t\d_\t+z\tb\d_{\tb}-2b\t\tb -4hz.
\ea\ee
The basis vectors of this super-vector space are
\be\ba{c}
\{<-2b,-2h|X_-,<-2b,-2h|(X_-)^2,\cdots, <-2b,-2h|(X_-)^{4h},\cr
<-2b,-2h|V_-,<-2b,-2h|X_-V_-,\cdots, <-2b,-2h|(X_-)^{4h-1}V_-,\cr
<-2b,-2h|W_-,<-2b,-2h|X_-W_-,\cdots, <-2b,-2h|(X_-)^{4h-1}W_-,\cr
<-2b,-2h|(V_-W_-),<-2b,-2h|(V_-W_-)^2,\cdots, <-2b,-2h|(V_-W_-)^{4h-2}\}.
\ea\ee
with dimension of $16h$.

Geometrically, this realization describes the left action of the group on
the sections of a holomorphic line bundle over the flag manifold $\os /T$,
where $T$ is maximal isotropic sub-algebra corresponding to the state
$<-2b,-2h\mid$ and $\mid 2b,2h>$,
$\{H,B\}$. $z$ and $\t$ are coordinates on the $\os /T$ and $h$ and $\b$ are the
coordinates on $T$.

For right action we choose the highest weight such that
\be\ba{cc}
X_+|2b,2h>=0,&V_+|2b,2h>=0,\cr
H|2b,2h>=(2h)|2b,2h>,&W_+|2b,2h>=0,\cr
B|2b,2h>=(2b)|2b,2h>.&
\ea\ee
The coherent state is
\be
e^{(z-\ff\t\tb)X_-}e^{\tb W_-}e^{\t V_-}|2b,2h>:=|z,\t,\tb>.
\ee
The relevant differential realization will be as follows:
\be\ba{c}
V_-=\d_{\tb}+\ff\t\d_z\cr
W_-=\d_{\t}+\ff\tb\d_z\cr
V_+=-z\d_{\tb}-\ff z\t\d_z-\ff\t\tb\d_{\tb}+2\t(h-b)\cr
W_+=-z\d_{\t}-\ff z\tb\d_z+\ff\t\tb\d_{\t}+2\tb (h+b)\cr
H=-\ff\t\d_\t-\ff\tb\d_{\tb}-z\d_z+2h\cr
B=-\ff\t\d_\t+\ff\tb\d_{\tb}+2b\cr
X_-=\d_z\cr
X_+=-z^2\d_z-z\t\d_\t-z\tb\d_{\tb}+2b\t\tb+4hz.
\ea\ee

where $\d_z=\f{\d}{\d z}$, $\d_\t=\f{\d}{\d_\t}$, $\d_\tb=\f{\d}{\d_\tb}$,
$[\d_z,z]=1 $, $\{\d_\t,\t\}=1$, $\{\d_\tb,\tb\}=1$, $\{\t,\tb\}=0$ and
$\{\d_\t,\d_\t\}=0$.

 One can consider
this, as the representation
of $\os$ on the super-sub-space of analytic functions spanned by the monomial,
$$\{1,z,z^2,...,z^{4h}, \t, \t z,\cdots, \t z^{4h-2}, \tb, \tb z,\cdots,
\tb z^{4h-2}, \t\tb, \t\tb z,\cdots, \t\tb z^{4h-4}\}$$
whose dimension is $16h$.
We will find that
this new operator realization,
which satisfies the algebra of (\ref{osc}),
is a finite-dimensional irreducible representation of $\os$.

Up to now, everywhere the covariant derivative in super-space was being defined
\cite{dgg} by
\be\ba{c}
D_{\t}:=\d_\t+\ff\tb\d_z\cr
D_{\tb}:=\d_\tb+\ff\t\d_z
\ea\ee
As we can see this is not a definition but it represents the differential
realization of $V_-$ and $W_-$ respectively on super-space of analytical
functions of $f(z,\zb,\t,\tb)$.
{\Section {Correlation functions of N=2 super-conformal field theory in d=2}}
\subsection {Two point correlation functions}
One of the simplest representations involves a complex scalar field $\p$ and
a two-component spinor $f_\a$ ($\a=1,2$). They form the so-called chiral 
multiplet.
One also should include a field $K$ that will turn out to be an auxiliary
field. Such a chiral super-field in terms of $z$ and $\zb$ as complex variables
and $\t$ and $\tb$ as anti-commuting variables is as follows:
\be\label{csf}
\Phi(z,\t)=\phi(z,\zb)+\t f(z,\zb)+ \tb \fb(z,\zb)+\t\tb K(z,\zb).
\ee
For the ease of notation we have abbreviated $(z,\zb,\t,\tb)$ to $(z,\t)$ on 
the left hand side.

To construct Ward identity according to the method which has been
given in \cite{wa}, we take following "co-product"
and differential realization of (\ref{lrz}), to calculate
the two point function,

\be
\Delta(g)=g\tp I+I\tp g;\hspace{0.5cm} g\in osp(2|2)
\ee
where $I$ means the identity operator.
Then the Ward identity is given by
\be
\begin{array}{c}
\Delta H\PP=(\s^{2}_{i=1}(\ff\t_i\d_{\t_i}+\ff\tb_i\d_{\tb_i}+z_i\d_{z_i}-2h_i))
)\PP=0\cr
\Delta B\PP=(\s^2_{i=1}(\ff\t_i\d_{\t_i}-\ff\tb_i\d_{\tb_i}-2b_i))\PP\cr
\Delta X_{-}\PP=(\s^{2}_{i=1}(\d_{z_i}))\PP=0\cr
\Delta V_{-}\PP=(\s^{2}_{i=1}(\d_{\tb_i} +\ff\t_i\d_{z_i}))\PP=0\cr
\Delta W_{-}\PP=(\s^{2}_{i=1}(\d_{\t_i} +\ff\tb_i\d_{z_i}))\PP=0\cr
\Delta V_+\PP=
(\s^2_{i=1}(z_i\d_{\t_i} +\t_i z_i\d_{z_i} +2h_i\t_i))
\PP=0\cr
\Delta X_+\PP=(\s^2_{i=1}\d_{z_i})\PP=0\cr
\end{array}
\ee
where $h_1$ and $h_2$ are conformal weight of $\Phi(z_1,\t_1)$ and
$\Phi(z_2,\t_2)$ respectively.

By solving the above Ward identities for super-fields, $\PP$, and using the
realization given by (\ref{lrz}), we have the following
expression for the two point function which is well-known \cite{q}
\be
\PP=(z_{12}-\f{1}{2}(\t_1\tb_2+\tb_1\t_2))^{\L}.
\ee
Our second attempt is to solve the above set of equations for two point 
functions of bosonic and spinor fields. The first equation in the above 
set will give the answer up to a constant. The second equation will fix 
the representation of the symmetric group $\os$, which here for chiral 
super-field, (\ref{csf}), the representation is $(2h,2b=0)$. The rest 
of equations will give the relations between different two point
functions.
\be\ba{l}
<\phi(z_1)\phi(z_2)>\sim z^\L\cr
<f(z_1)\fb(z_2)>=<\fb(z_1)f(z_2)>\sim\f{1}{2}\L z^{(\L-1)}\cr
<K(z_1)K(z_2)>\sim\f{1}{2}\L(\L-1)z^{(\L-2)}\cr
\ea\ee
where
$$
z=(z_1-z_2),\hspace{1cm}\L=2(h_1+h_2)$$
Some remarks are in order:\\
1- Correlation functions are scaling.\\
2- A specific representation of the algebra has been chosen.\\
These two point are the indications that such a model is a non-unitary model
\cite{mas}.

\subsection {Three point functions}
A method similar to that in last section can be used to calculate three point correlation
functions.
For three point function the result is as follows:
\be
\PPP\sim {z_{12}^{(h_1+h_2-h_3)}z_{13}^{(h_1-h_2+h_3)}
z_{23}^{(-h_1+h_2+h_3)}}(1+\beta\hat{\eta}),
\ee
where $\beta$ is constant,
$$
\hat{\eta}=\f{\t_1z_{23}+\t_2z_{13}+\t_3z_{12}}{\sqrt{z_{12}z_{13}z_{23}}},
\hspace{0.25cm}z_{ij}=(z_i-z_j-\t_i\t_j).$$
\be\ba{l}
<\phi(z_1)\phi(z_2)\phi(z_3)>\sim z_{12}^{2\a}z_{13}^{2\b}z_{23}^{2\g}\cr
<\phi(z_1)f(z_2)\fb(z_3)>\sim\ff\g z_{12}^{2\a}z_{13}^{2\b}z_{23}^{2\g-1}\cr
<f(z_1)\phi(z_2)\fb(z_3)>\sim\ff\b z_{12}^{2\a}z_{13}^{2\b-1}z_{23}^{2\g}\cr
<f(z_1)\fb(z_2)\phi(z_3)>\sim\ff\a z_{12}^{2\a-1}z_{13}^{2\b}z_{23}^{2\g}\cr
<\phi(z_1)K(z_2)K(z_3)>\sim\ff\g(\g-1)z_{12}^{2\a}z_{13}^{2\b}z_{23}^{2\g-2}\cr
<K(z_1)\phi(z_2)K(z_3)>\sim\ff\b(\b-1)z_{12}^{2\a}z_{13}^{2\b-2}z_{23}^{2\g}\cr
<K(z_1)K(z_2)\phi(z_3)>\sim\ff\a(\a-1)z_{12}^{2\a-2}z_{13}^{2\b}z_{23}^{2\g}\cr
<f(z_1)\fb(z_2)K(z_3)>\sim\ff\b\g z_{12}^{2\a}z_{13}^{2\b-1}z_{23}^{2\g-2}\cr
<K(z_1)f(z_2)\fb(z_3)>\sim\ff\a\b z_{12}^{2\a-1}z_{13}^{2\b-1}z_{23}^{2\g}\cr
<f(z_1)K(z_2)\fb(z_3)>\sim\ff\a\g z_{12}^{2\a-1}z_{13}^{2\b}z_{23}^{2\g-1}\cr
<K(z_1)K(z_2)K(z_3)>\sim\a\b\g z_{12}^{2\a-1}z_{13}^{2\b-1}z_{23}^{2\g-1}\cr
\ea\ee
and the rest are zero. Here,
$$
z_{ij}=z_i-z_j,\hs{0.5cm} \a=h_1+h_2-h_3,\hs{0.5cm}\b=h_1-h_2+h_3,\hs{0.5cm}
\g=-h_1+h_2+h_3.$$
In the case of three point functions also, we find out that the representation,
$(2h,0)$, is the only one valid for the super-field given in (\ref{csf}) and
all other correlations are zero.

{\Section {Free Field Representation of ${\widehat{\os_k}}$}}
In parallel to Wakimoto's method \cite{wak} of free field realization
of given current algebra, $ \hat{\cal G}$, we use the simple observation
that at the zero level these currents, ${\hat g}\in{\hat{\cal G}}$,
become differential operator realization of ${\cal G}$ given by (\ref{lrz}).
By using the Feigin and Frenkel \cite{fefr} method we can extend level zero to
any arbitrary level $k$.

To get the free field realization of currents in standard form, we the
make following changes:
\be\ba{c}
z\rar \chi,\hs{0.5 cm}\d_z\rar W,\hs{0.5 cm}2h\rar \sqrt{2(k-1)}\d\phi:=j_0,
\t\rar\psi_-,\cr
\d_\t\rar\psi_+,\hs{0.5 cm}\tb\rar{\bar{\psi}_-},\hs{0.5 cm}
\d_\tb\rar{\bar{\psi}_+},\hs{0.5 cm} 2b\rar 2\d\phi':=j_0'.
\ea\ee
According to section (2), the classical part of affine algebra $G$ (zero level
affine algebra ${\hat{\cal{ G}}}$) in fact comes from the action of $G$ on
homogenous space (flag manifold) $G/T$ as an algebra of vector fields.
Then $\chi_i$'s and $\psi$'s are
the corresponding complex and Grassmanian coordinates on $G/T$ 
respectively, and $W_i\sim\f{\d}{\d\chi_i}$.
The fields $\phi$ and $\phi'$ are the corresponding coordinates on $T$.
The Killing vectors
$J(W, \chi, \psi, \bar{\psi})$ of zero level affine algebra ${\hat{\cal G}}$
do not depend on
$\phi$ and $\phi'$, but such decoupling no longer takes place when
$W$, $\chi$, $\psi$, $\bar{\psi}$, $\phi$ and
$\phi'$ are considered $z$-dependent, and affine algebra
${\widehat G}$ arises instead of classical finite-dimensional $G$.
So, we considered only flag manifolds with $T$ being a product of $U(1)$
factors, and this provided us with the free field realization of the model.

As a result the free field realization of ${\hat{\cal G}}$ generators will be
\be\ba{c}\label{ffk}
V_-={\bar{\psi_+}}+\ff\psi_-W\cr
W_-=\psi_++\ff\bar{\psi}_-W\cr
V_+=\chi\bar{\psi}_++\ff \chi\psi_-W+\ff\psi_-\bar{\psi}_-\bar{\psi}_+
    -\psi_-(\a j_0-\a'j_0')+\b\d\psi_-\cr
W_+=\chi\psi_++\ff \chi{\bar{\psi}_-}W-\ff\psi_-
   {\bar{\psi}_-}{\bar{\psi}_+}-\psi_-(\a j_0+\a'j_0')+\b'\d\bar{\psi}_-\cr
H=\ff\psi_-\psi_++\ff{\bar{\psi}}_-{\bar{\psi}}_++\chi W-\a j_0\cr
B=\ff\psi_-\psi_+-\ff\bar{\psi}_-\bar{\psi}_+-\a'j_0'\cr
X_-=W\cr
X_+=\chi^2W+\chi\psi_-\psi_++\chi\bar{\psi}_-\bar{\psi}_+
-\a'j_0'\psi_-\bar{\psi}_--2\a\chi j_0+\g\d\chi.
\ea\ee
The constituents are free fields whose commutation relations are encoded by
the singular OPEs
\be\ba{cc}
W(z)\chi(\o)\sim\f{-1}{z-\o},& \d\phi(z)\d\phi(\o)\sim\f{-1}{(z-\o)^2},\cr
\psi^i_-(z)\psi^i_+(\o)\sim\f{1}{z-\o};&
\psi^1=\psi,\;\psi^2=\bar\psi.
\ea\ee
The non-trivial OPE of the above currents will be as follow:
\be\ba{ll}
H(z)H(\o)\sim\f{-(\ff+\a^2)}{(z-\o)^2}+\cdots&
B(z)B(\o)\sim\f{\ff}{(z-\o)^2}+\cdots\cr
H(z)V_{\pm}(\o)\sim\f{\pm\ff V_\pm}{(z-\o)}+\cdots&
H(z)W_{\pm}(\o)\sim\f{\pm\ff W_\pm}{(z-\o)}+\cdots\cr
B(z)V_{\pm}(\o)\sim\f{\ff V_\pm}{(z-\o)}+\cdots&
B(z)W_{\pm}(\o)\sim\f{-\ff W_\pm}{(z-\o)}+\cdots\cr
H(z)X_{\pm}(\o)\sim\f{\pm X_\pm}{(z-\o)}+\cdots&
V_\pm(z)W_\mp(\o)\sim\f{H\pm B}{(z-\o)^2}+\f{\mp(\b+\ff)}{(z-\o)^2}+\cdots\cr
X_+(z)X_-(\o)\sim\f{\g}{(z-\o)^2}-\f{2H}{(z-\o)}+\cdots&
V_\pm(z)W_\pm(\o)\sim\f{X_\pm}{(z-\o)}+\cdots\cr
X_\pm(z)V_\mp(\o)\sim\f{\mp V_\pm}{(z-\o)}+\cdots&
X_\pm(z)W_{\pm}(\o)\sim\f{\mp W_\pm}{(z-\o)}+\cdots
\ea\ee
where $\cdots$ are non-singular terms.
In calculation of the above OPEs we found the system
of equations for $\a$, $\a'$, $\b$ and $\g$, which result in
$$\a'=\a=0,\hs{0.5cm}\b=\ff,\hs{0.5cm}\g=1.$$

Such a result shows that
the representation given in (\ref{ffk}) is free field representation of
$\widehat{\os}$ in level one (k=1).
The above expression can be put in the following compact form:
\be
J^a(z)J^b(\o)\sim k\f{\kappa^{ab}}{(z-\o)^2}+{f^{ab}}_c\f{J^c(\o)}{z-\o}.
\ee
where ${f^{ab}}_c$ are the structure constants of $\os$, and $\kappa^{ab}$ is
proportional to its non-degenerate Killing form.

The stress-energy tensor of the above model can be obtained by means of
the Sugawara construction,
$$
T(z)=\f{1}{\kappa}\kappa_{ab}:J^a(z)J^b(z):
$$
where $\kappa_{ab}\kappa^{ab}=\pm 1$, $+1$ for compact and $-1$ for non-compact
groups. The result is
\be\ba{l}
T(z)=3W(z)\d\chi(z)+\chi(z)W(z){\bar\psi}_-(z){\bar\psi}_+(z)
+\chi(z)W(z)\psi_-(z)\psi_+(z)
+{\bar\psi}_-(z)\d {\bar\psi}_+(z)\cr
+{\bar\psi}_+(z)\d{\bar\psi}_-(z)
+\psi_-(z)\d\psi_+(z)
-\psi_+(z)\d\psi_-(z)
+\psi_-(z)\psi_+(z){\bar\psi}_-(z){\bar\psi}_+(z)
\ea\ee
By using the OPE of stress-tensor
$$
T(z)T(\o)\sim\f{c/2}{(z-\o)^4}+2\f{T(\o)}{(z-\o)^2}+\f{\d T(\o)}{z-\o)}
$$
one can get the central charge, $c$, as zero. This result was evident from
$$
c=2k\f{\kappa_{ab}\kappa^{ab}}{\kappa}
$$
where $\kappa$ as a constant is such that $J^a(z)$ to be a primary
field of conformal weight one:
$$
T(z)J^a(\o)\sim\f{J^a(\o)}{(z-\o)^2}+\f{\d J^a(\o)}{z-\o}.
$$
Hence, the conformal field theory with $\os$ symmetry is a non-unitary
$c=0$ model.\\ \\
\noindent
{\bf Acknowledgements:}\\
The authors would like to thanks, M. Khorrami and A. Morozov
for their useful discussions.\\
The work of W.S.C. was supported by the KOSEF (961-0201-004-2)
and the present studies were supported by Basic Science
Research
Program, Ministry of Education of Korea, 1996(BSRI-94-2413).

\end{document}